\documentclass[aps,prb,superscriptaddress,twocolumn,floatfix,amsmath,amssymb]{revtex4-2}
\usepackage{graphicx}% Include figure files
\usepackage[
colorlinks, 
citecolor=blue, 
urlcolor=blue, 
linkcolor=red,breaklinks]{hyperref}

\begin{document}

\title{Flat band localization in twisted bilayer graphene nanoribbons}

\author{Elias Andrade}
\email{eandrade@estudiantes.fisica.unam.mx}
\affiliation{
 Posgrado en ciencias Físicas, Instituto de F\'{i}sica, Universidad Nacional Aut\'{o}noma de M\'{e}xico (UNAM). Apdo. Postal 20-364, 01000 M\'{e}xico D.F., M\'{e}xico
 }%

\author{Pierre A. Pantale\'on}
\affiliation{IMDEA Nanoscience, Faraday 9, 28049 Madrid, Spain}

\author{Francisco Guinea}
\affiliation{IMDEA Nanoscience, Faraday 9, 28049 Madrid, Spain}
\affiliation{Donostia International Physics Center, Paseo Manuel de Lardiz\'abal 4, 20018 San Sebastián, Spain}
\affiliation{ Ikerbasque, Basque Foundation for Science, 48009 Bilbao, Spain}

\author{ Gerardo G. Naumis}
\email{naumis@fisica.unam.mx}
\homepage{\\http://www.fisica.unam.mx/personales/naumis/}
\affiliation{
 Depto. de Sistemas Complejos, Instituto de F\'{i}sica, Universidad Nacional Aut\'{o}noma de M\'{e}xico (UNAM). Apdo. Postal 20-364, 01000 M\'{e}xico D.F., M\'{e}xico
 }%

\date{\today}

\begin{abstract}

We analyze the electronic structure of twisted bilayer graphene (TBG) nanoribbons close to the magic angle. We describe a transition from an incomplete to a complete moir\'e structure. By considering \textit{zigzag} and \textit{armchair} edge terminations, the low-energy bands are strongly modified, and thus, the edge flat band localization is sensitive to the type of boundary. By means of a scaled tight-binding model, we calculate the band structure and find that, for an \textit{armchair} configuration, an incomplete moir\'e edge suppresses the edge localization, while for a \textit{zigzag} configuration, we find a strong interference of the edge states with the moir\'e bands. In particular, for the \textit{armchair} termination, we observe a competition between the ribbon periodicity and the graphene monolayers, which we describe with a potential well toy model. Furthermore, for ribbons with widths of multiple moir\'e cells, the flat bands of the moir\'es in the bulk are unperturbed as we change the borders. These results are explained in terms of the strong electronic localization, nearly Gaussian, in the AA stacking regions, as confirmed by an inverse participation ratio analysis. Our results demonstrate that the electronic structure of TBG nanoribbons is sensitive to the edge termination, offering an explanation for recent experimental results.

\end{abstract}

\maketitle
\section{Introduction}
Due to the interplay between non-trivial band topology and band flatness, magic angle twisted bilayer graphene (TBG) has shown to be an ideal platform for the appearance of a plethora of topological and correlated phases, including for example non-conventional superconductivity \cite{cao2018unconventional,yankowitz2019tuning}, correlated insulators \cite{cao2018correlated,nuckolls2020strongly,pierce2021unconventional}, ferromagnetism \cite{sharpe2019emergent}, and other many body effects~\cite{lu2019superconductors,polshyn2019large,serlin2020intrinsic,chen2020tunable,saito2020independent,zondiner2020cascade,wong2020cascade,stepanov2020untying,xu2020correlated,choi2021correlation,rozen2021entropic,cao2021nematicity,stepanov2021competing,oh2021evidence,xie2021fractional,berdyugin2022out,turkel2022orderly,huang2022observation}. For magic angles the flat bands near the Fermi energy appear due to the localization of the electronic wavefunction at the $AA$-stacking sites, this was predicted~\cite{SanJose2012NonAbelian,DosSantos2012Continuum, Trambly2012Confined} and then experimentally confirmed~\cite{Luican2011SingleBreak,Experimental1,Experimental2,Experimental3,Experimental4}. The wavefunction geometry comes from the modulated interlayer coupling, which can be seen as an effective moir\'e potential well~\cite{LopesdosSantos2007,Shallcross2008TwistB, Shallcross2010Turbo,Bistritzer2011,TramblydeLaissardire2010,Mele2010Comm}.  

Studies of TBG with reduced translational symmetry, for example in flakes, have shown that a single moir\'e cell may be enough to localize the wavefunction in the $AA$ spots~\cite{Landgraf2013FlakeEdges}, supporting a  \textquotedblleft moir\'e quantum well\textquotedblright~picture. However, theoretical~\cite{Morell2014TBGnanorribbons,Morell2015EdgeGraphite,Shao2022EdgeStates,fleischmann2018moire} and experimental~\cite{FortinDeschnes2022,Wang2023ZigZagNano} studies have demonstrated that both bulk and \textquotedblleft moir\'e edge states\textquotedblright~appear in TBG with a nanoribbon geometry. In particular, it has been shown that in the presence of a \textit{zigzag} termination there is a coexistence of edge and bulk moir\'e states near the Fermi energy~\cite{Wang2023ZigZagNano}, and that the edge states are found at the edges of $AB$ sites~\cite{Morell2014TBGnanorribbons,Morell2015EdgeGraphite}. 

On the other hand, recent experiments have shown the breakdown of moir\'e flat bands in TBG due to edge termination~\cite{yin2022direct}.  In the experiment, the electronic localization is found to be persistent even if translational symmetry is broken as long as the moir\'e supercell is complete. However, when the supercell is incomplete, there is a strong suppression of the electronic localization resulting in a vanishing of the density of states near the Fermi Energy. 

A useful methodology to analyze the transition from a complete to an incomplete supercell is by introducing an interlayer sliding between graphene layers, where the displacement between layers is particularly relevant in terminated TBG due the change in the borders at both the scales of the moir\'e lattice and the graphene lattice. This type of displacement is known to produce a topological charge pumping in the perpendicular direction of the displacement, characterized by a topological invariant known as the sliding Chern number~\cite{Ying2020pumping,Zhang2020pumping,Fujimoto2020edgepumping,Fujimoto2021pumping}. 

In spite of this, while the topological sliding properties provide a description of the number of moir\'e edge states, the mechanism behind the breakdown of the flat bands as the portion of the moir\'e cell is reduced at the edges, as well as its relationship with electronic localization, is not clear. To address these questions, in this work, we analyze the electronic properties of TBG nanoribbons with different unit cells, ranging from an incomplete to a complete moir\'e supercell. We consider two nanorribon geometries: \textit{zigzag} and \textit{armchair}~\cite{Fujita1996Peculiar,Nakada1996EdgeState}. For each configuration, we monitor the electronic localization and the group velocity of the bands near the Fermi energy, and we distinguish between the contributions originating from the moir\'e scales of those of the graphene lattice. We find that the electronic structure is strongly sensitive to the edge configuration which may lead to the coexistence or suppression of edge and bulk moir\'e electronic localization.  Our work demonstrates the significant dependence of the electronic structure of twisted bilayer graphene nanorribbons with the edge termination and may offer an explanation for the breakdown of the moir\'e flat bands found in Ref.~\cite{yin2022direct}.  

The layout of this paper is as follows. In Sec.~\ref{sec:Model}, we present the model used composed of a minimum tight binding Hamiltonian~\cite{lin2018minimum} plus a rescaling approximation. In Sec.~\ref{sec:AC} we present our main numerical results concerning \textit{armchair} terminated TBG nanoribbons, where we analyze the change on the electronic properties as a function of the completeness of the moir\'e supercell within the ribbon.  In Sec.~\ref{sec:toymodel} we discus the change in energy of the bands as we increase the size of the moir\'e and explain them qualitatively through a toy-model. We also include a brief analysis regarding \textit{zigzag} terminated TBG nanoribbons in Sec.~\ref{sec:zz}. Finally in Sec.~\ref{sec:conclusion} we give our conclusions.

\section{Model}
\label{sec:Model}
\begin{figure}
\begin{center}
\includegraphics[width=0.45\textwidth]{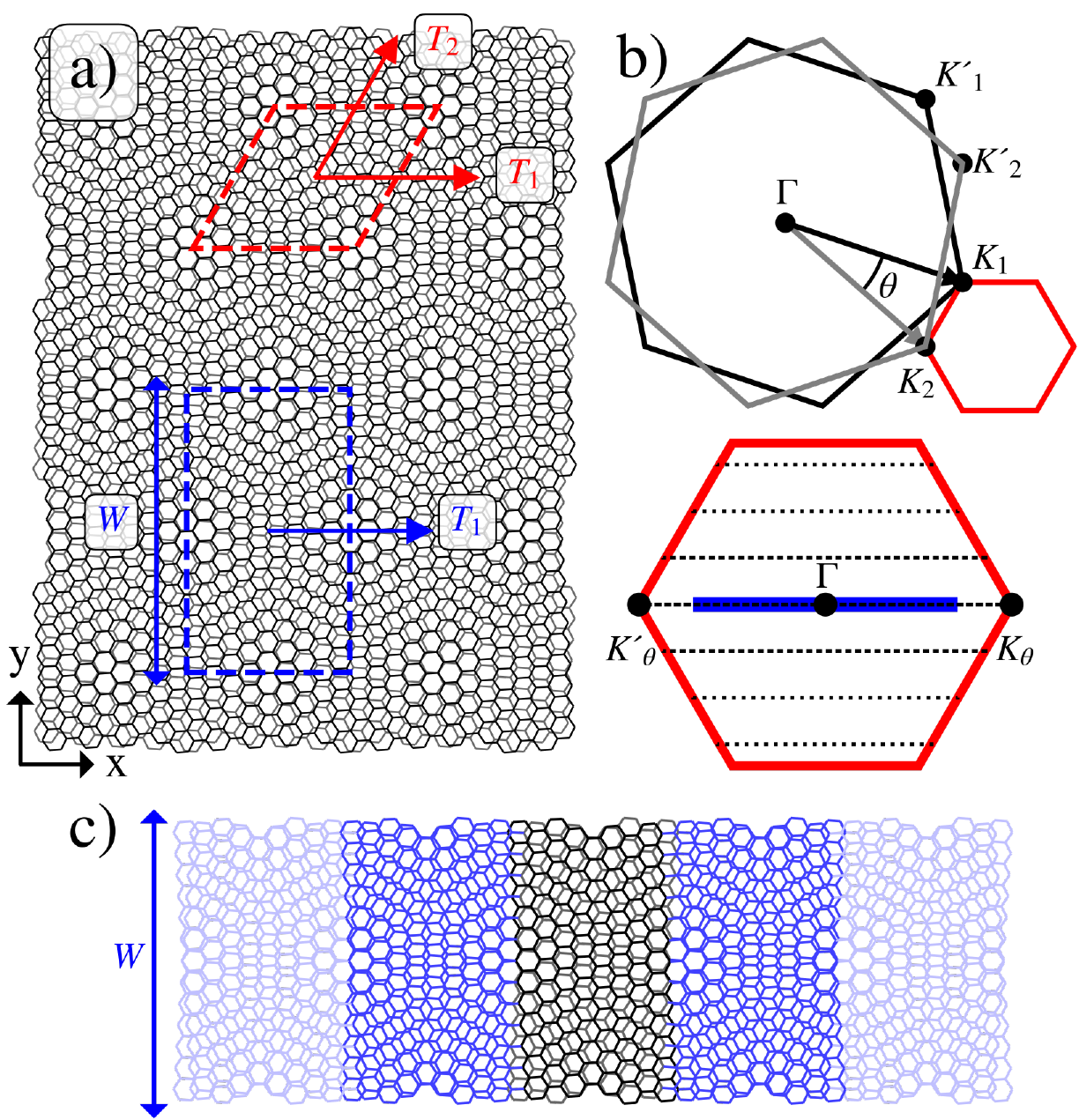}
\caption{ a) Lattice structure of TBG  with $\theta=7.34^o$, red dashed line is the corresponding unit cell. The blue square cell is used to construct a nanoribbon of width $W$. b) Mini-Brillouin zone (red) resulting from the twist between layers. For the nanoribbon the Brillouin gets folded into the blue line as the momentum along the $y$-direction is quantized. c) Formation of the nanoribbon with the unit cell shown in a), here five unit cells are shown.}
\label{fig:system}
\end{center}
\end{figure}
We consider a TBG lattice as shown in Fig.~\ref{fig:system}a). The structure is defined by rotating the layer 1 and 2 of the AA-stacked bilayer around the common center by $\pm \theta/2$, respectively. The lattice vectors of graphene are $a_1=\sqrt{3}a(0,1)$ and $a_2=\sqrt{3}a(\frac{\sqrt{3}}{2},\frac{1}{2})$, where $a \approx 1.42$~\text{\AA} is the distance between carbon atoms. Upon rotation, the lattice vectors on each layer are modified as $a_{1,2}^{u/d}=R(\pm \theta/2)a_{1,2}$. We consider a commensurate lattice~\cite{Conmensurate1,Conmensurate2} such that the mismatch between layer produces a moir\'e pattern with lattice vectors given by
\begin{equation}
    T_{1/2}=\frac{1}{2\text{sin}(\theta/2)}R(-30^o)a_{1/2}.
\end{equation}
We employ a minimum tight-binding Hamiltonian \cite{lin2018minimum}, which consists of an intralayer part $H_{||}$ and an interlayer part $H_{\perp}$, given by,
\begin{equation}
\begin{split}
    H&=H_{||}+H_{\perp}\\
    &=-\sum_{i\neq j,m}\gamma_{ij}^{m,m}(c_{m,i}^{\dagger}c_{m,j}+h.c.)\\
    &\quad -\sum_{i,j,m}\gamma_{ij}^{m,m+1}(c_{m,i}^{\dagger}c_{m+1,j}+h.c.),
\end{split}
\label{Eq:Ham}
\end{equation}
where $c_{m,i}^{\dagger}$ and $c_{m,i}$ are creation and annihilation operators respectively, acting on site $i$ and layer $m$, $\gamma_{ij}^{m,m}$ and $\gamma_{ij}^{m,m+1}$ are the intralayer and interlayer hopping integrals, respectively.
For the intralayer hoppings only first-neighbors are considered, which have a constant value $\gamma_{ij}^{m,m}=t_{||}=3.09$ eV. The interlayer hopping is considered to decay exponentially as,
\begin{equation}
    \gamma_{ij}^{m,m+1}=t_{\perp}\frac{d_0^2}{r^2+d_0^2}\text{exp}\left(-\frac{\sqrt{r^2+d_0^2}-d_0}{\lambda_{\perp}}\right),
\end{equation}
where $t_{\perp}=0.39$ eV is the hopping amplitude, $d_0=3.35$ \text{\AA} is the distance between layers and $\lambda_{\perp}=0.27$ \text{\AA} is a modulation of the interlayer hopping. 

To reduce the number of atoms and the computational cost in the numerical calculations, we employ a re-scaling approximation~\cite{ScalingGuinea,ScalingVahedi,ScalingNanotubes}, where the bands in TBG depend only on a single dimensionless parameter $\alpha$ given by~\cite{bistritzer2011moire},

\begin{equation}
    \alpha=\frac{\sqrt{3}at_{\perp}}{2 \hbar v_f \text{sin} (\theta/2)}\propto \frac{t_{\perp}}{t_{||}\text{sin}(\theta/2)},
\end{equation}
here $t_{\perp}$ refers to an average of the interlayer hopping and $v_f$ is the Fermi velocity. The importance of this relation is that the dependence on the ratio of the hopping integrals and angle allows us to explore the low-energy physics of a TBG system with a given $\alpha$ using a larger angle and thus reducing the number of atoms in the unit cell. To do this, we use re-scaled intralayer hopping integral,
\begin{subequations}
\begin{equation}
    t_{||} \rightarrow \frac{t_{||}}{\Lambda },
\end{equation}
where,
\begin{equation}
    \Lambda=\frac{\text{sin}(\theta'/2)}{\text{sin}(\theta/2)}.
\end{equation}
\label{eq:scaling}
\end{subequations}
In the above equation, $\theta'$ is the angle used to simulate the spectrum of a system with angle $\theta$. Furthermore, to preserve the length of the moir\'e-cell, a re-scaling of the distances must be done as well,
\begin{equation}
    a \rightarrow \Lambda a.
\end{equation}
This re-scaling approximation allow us to simulate the magic angle $\theta=1.08^o$ which has a unit cell of about $11\times10^3$ atomic sites. 
In particular, with an angle $\theta'=2.28^\circ$ which has 5048 atoms, the low energy band structure is nearly identical to that of the magic angle if we use an scaling factor of $\Lambda=2.11$ in Eq.~\ref{eq:scaling}, thus making computational calculations more attainable. 

\section{Armchair Terminated TBG Nanoribbons} \label{sec:AC}

\subsection{Electronic Structure}
\begin{figure}
\begin{center}
\includegraphics[width=0.48\textwidth]{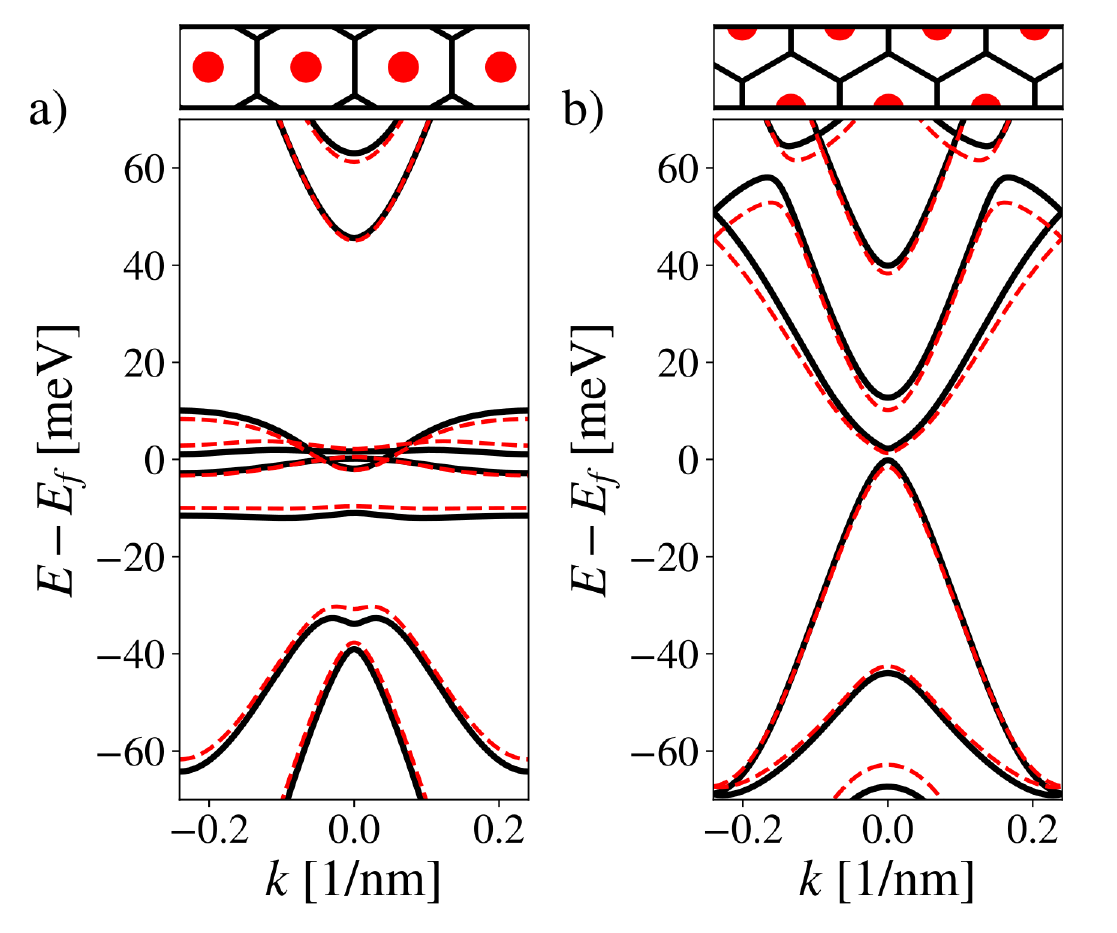}
\caption{Low energy band structure for a TBG nanorribon of the width of one moir\'e cell with twist angle $\theta=1.08^o$ (black solid line) and with $\theta'=2.28^o$ using the rescaling to simulate $\theta$ (red dashed line). Above each band structure a schematic of the geometry of each ribbon is shown, where the red circles indicate the $AA$ stacking regions. For a) the moir\'e is complete with the $AA$ regions located at the center of the ribbon. As the moir\'e cell is complete, states get localized here and flat bands appear around the Fermi energy. In b) the moir\'es are split in halves with the $AA$ regions at the edges, the spectrum around the Fermi energy is dispersive.}
\label{fig:Rescaling}
\end{center}
\end{figure}

By using the methodology described in the previous section, we first consider the case of \textit{armchair} terminated TBG nanoribbons. This configuration is the same as that in Fig.~\ref{fig:system}c) where the periodicity is along the horizontal direction. In Fig.~\ref{fig:Rescaling} we show the corresponding spectrum with a single moir\'e $\approx 11.3$ nm considering different moir\'e terminations. In Fig. \ref{fig:Rescaling}a) the $AA$ regions at the moir\'e center coincide with the ribbon center and the moir\'e cell is mostly complete, thus the usual localization at the $AA$ regions occurs and flat bands can be seen, in agreement with previous results in TBG flakes~\cite{Landgraf2013FlakeEdges}. Meanwhile, in Fig. \ref{fig:Rescaling}b) the moir\'e cells are split in halves with the $AA$ regions at the edges. These half-moir\'e cells are not sufficient to localize the wavefunction at low-energies and the resulting spectrum is dispersive.

\begin{figure*}
\begin{center}
\includegraphics[width=0.98\textwidth]{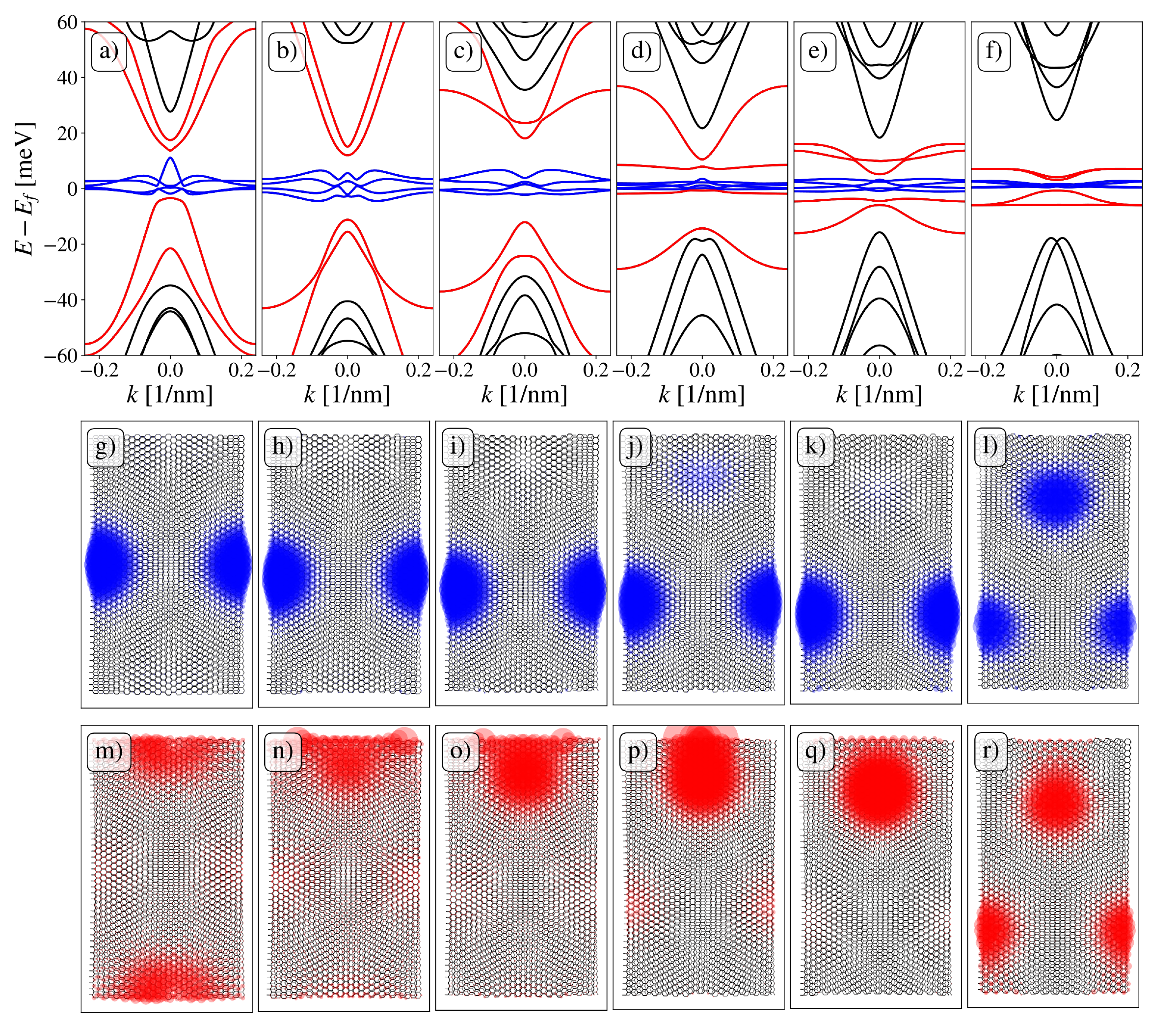}
\caption{Low energy band structure for {\it armchair} terminated TBG nanoribbons  simulating a twist angle $\theta=1.08^o$ with constant width $W=22.6$ nm, we consider different moir\'e terminations by translating the unit cell of the ribbon by a) $\delta=0$, b) $\delta=0.05W$, c) $\delta=0.1W$, d) $\delta=0.15W$, e) $\delta=0.2W$, f) $\delta=0.25W$. The blue bands are the four lowest energy bands, these states are  localized at the complete moir\'e. The next four energy bands are shown in red, these correspond to the incomplete moir\'e and as we change the termination, increasing the size of the incomplete moir\'e the bands get flatter and go towards $E_f$. In g)-l) we show the total charge density of the blue bands respectively to a)-f) within the unit cell, while in m)-r) for the red bands.}
\label{fig:ACRibbons}
\end{center}
\end{figure*}

In Fig.~\ref{fig:ACRibbons} we further study this transition in ribbons of width $W=22.6$ nm, around the length of two moir\'es, ensuring that there is always at least one complete moir\'e within the ribbon. We characterize different moir\'e terminations with a sliding parameter $\delta$~\cite{Ying2020pumping,Zhang2020pumping,Fujimoto2020edgepumping,Fujimoto2021pumping}, such that it translates the ribbon cell within the TBG lattice, starting from $\delta=0$, cf. Fig.~\ref{fig:ACRibbons}g), where the moir\'e is centered with the ribbon, to $\delta=0.25W$, cf. Fig.~\ref{fig:ACRibbons}l), where the center of the ribbon is between two moir\'es. In Fig.~\ref{fig:ACRibbons}a)-f) we show the energy spectrum around the Fermi energy ($E_f$) for six different configurations. For each band structure, we show in blue the four lowest energy bands and the next four in red. Note that the sliding structure is periodic in $\delta$ such that $\delta = 0.3W$ corresponds to Fig.~\ref{fig:ACRibbons}e) with the unit cell upside down. In Figs.~\ref{fig:ACRibbons}g)-l) and  Figs.~\ref{fig:ACRibbons}m)-r) we show the corresponding total charge density of the bands for each system.

\begin{figure}[!htbp]
\begin{center}
\includegraphics[width=0.48\textwidth]{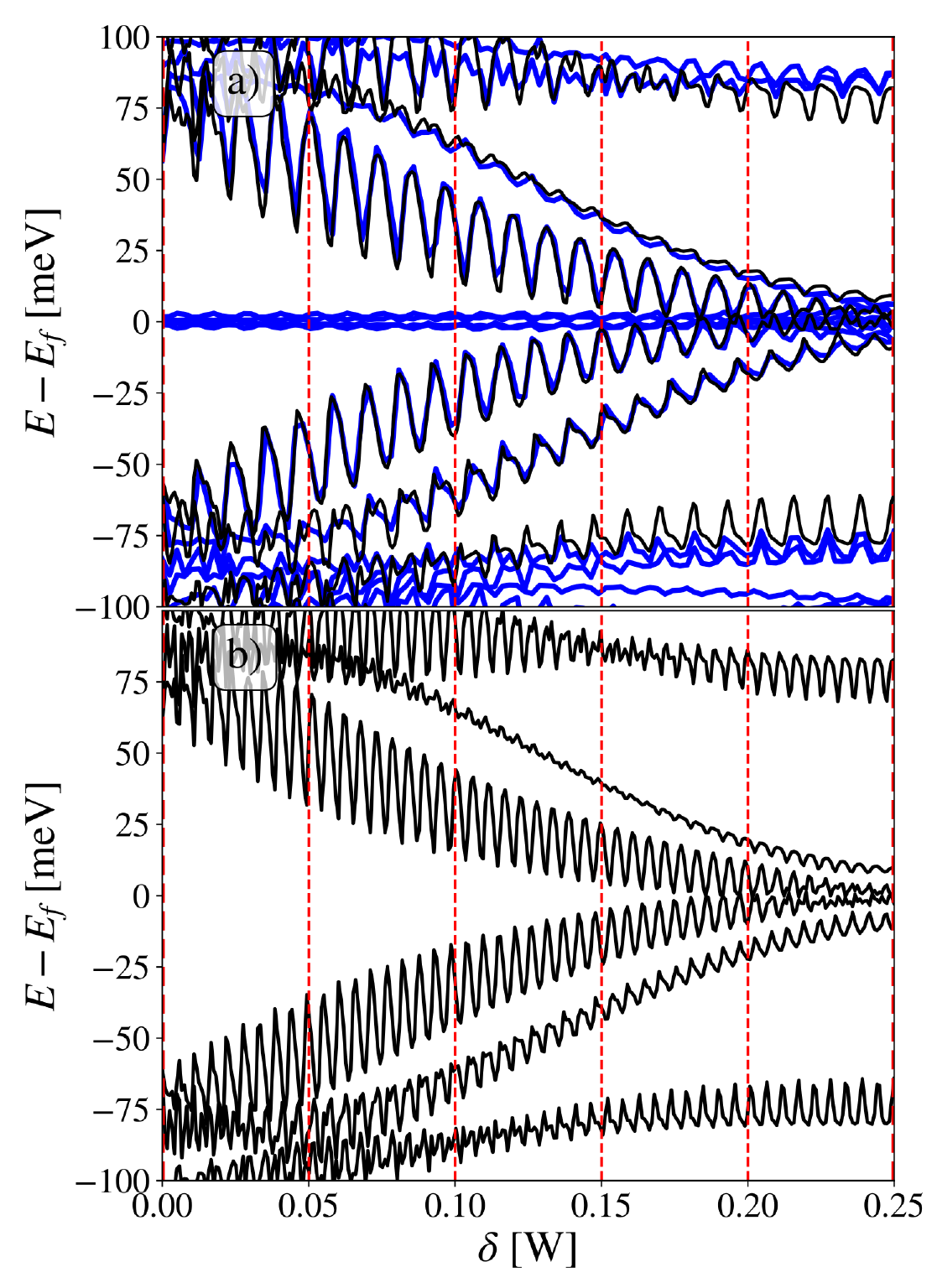}
\caption{Spectrum at the edge of the Brillouin zone as a function of $\delta$. For a) the rescaling is used, we show in black the spectrum for a ribbon of the width of one moir\'e and in blue for a width of two moir\'es. It can be seen how the four bands go towards $E_f$ as we increase $\delta$, furthermore these bands lower their energy independently of the width of the ribbons, as these bands only depend on the moir\'e that is being formed at the border. The complete moir\'es in the bulk will just result in more flat bands around $E_f$. In b) for the width of one moir\'e without rescaling, the wavelength of the oscillation increases as these come from the microscopic borders which is not equal within the re-scaling, but the overall behavior is retained. The dashed red vertical lines indicate the values of $\delta$ used in Fig. \ref{fig:ACRibbons}.}
\label{fig:MoireEdges}
\end{center}
\end{figure}

In Fig.~\ref{fig:ACRibbons}a) for $\delta =0$, there are four narrow bands near $E_f$ which correspond to states mainly localized at the $AA$ regions within the whole moir\'e cells. There is also localization at the incomplete moir\'es, but it occurs at higher energies, as shown in Fig.~\ref{fig:ACRibbons}m). Interestingly, by following the sequence of panels in Fig.~\ref{fig:ACRibbons}, as we increase $\delta$, the half moir\'e cell at the bottom starts to disappear while the upper one grows. Intuitively, the increase in the upper portion of the moir\'e allows the localization of the wavefunction in a greater area, lowering its energy towards $E_f$, and also reducing its group velocity. This effect is clearly seen in  Figs.~\ref{fig:ACRibbons}a)-d) where four red dispersive bands are flattened and pushed towards the center of the energy spectrum, thus increasing their localization in the AA centers, as shown in Figs.~\ref{fig:ACRibbons}g)-r). 

However, although there is an increase in the size of the incomplete moir\'e, the red band closer to $E_f$ in Fig.~\ref{fig:ACRibbons}d), is further away in Fig.~\ref{fig:ACRibbons}e), which is also reflected in the absence of hybridization between states of the two moir\'es in Fig.~\ref{fig:ACRibbons}k) and Fig.~\ref{fig:ACRibbons}q). This indicates that the decreasing of the moir\'e bands energy towards $E_f$ is non-monotonous.  We further explore this behavior in Fig.~\ref{fig:MoireEdges}, where we plot the spectrum at the boundary of the Brillouin zone as a function of $\delta$. We observe an oscillatory behavior of the energy bands, which we found to be influenced by the sliding parameter in two distinct ways. Firstly, there is a general tendency of the remote bands to transition from being dispersive to becoming narrow and merge with the flat bands. This occurs near $\delta=0.25W$. By continuing to increase the sliding parameter, it is equivalent to moving backward in the plot. This large-scale wavelength oscillation has been described as a topological sliding effect in Refs.~\cite{Ying2020pumping,Zhang2020pumping,Fujimoto2020edgepumping,Fujimoto2021pumping}. Secondly, the bands display a short wavelength oscillation not described before, which we attribute to an interference effect at the atomic scale, (cf. Sec.~\ref{sec:toymodel}).  

To analyze these oscillatory effects, in Fig.~\ref{fig:MoireEdges}a), we show the bands for two different ribbon widths (black and blue lines), while Fig.~\ref{fig:MoireEdges}b) displays the same bands without the scaling approximation. Thus, a comparison between both figures indicates that the long wavelength oscillations remain unchanged, but the short wavelength oscillations are different. This suggests that the transition of the bands from being dispersive to becoming narrow is a moir\'e scale effect, is independent of the ribbon size, and is mainly controlled by the percentage of the incomplete moir\'e at the boundary. Furthermore, we notice that without the re-scaling of the atomic distances, the short wavelength oscillation in Fig.~\ref{fig:MoireEdges}b) is invariant, and has a value of $\lambda=\sqrt{3}a/2$. This associate the oscillatory effect to the edges of the microscopic graphene lattice and cannot be effectively mapped with the re-scaling approximation. Thus, in Fig.~\ref{fig:MoireEdges}a), due to the re-scaling, the oscillation has a larger wavelength given by $\lambda'=\Lambda \lambda$ in comparison with Fig.~\ref{fig:MoireEdges}b) without re-scaling. We further describe this effect in Sec.~\ref{sec:toymodel}.

\begin{figure}%[h]%[!htbp]
\begin{center}
\includegraphics[width=0.48\textwidth]{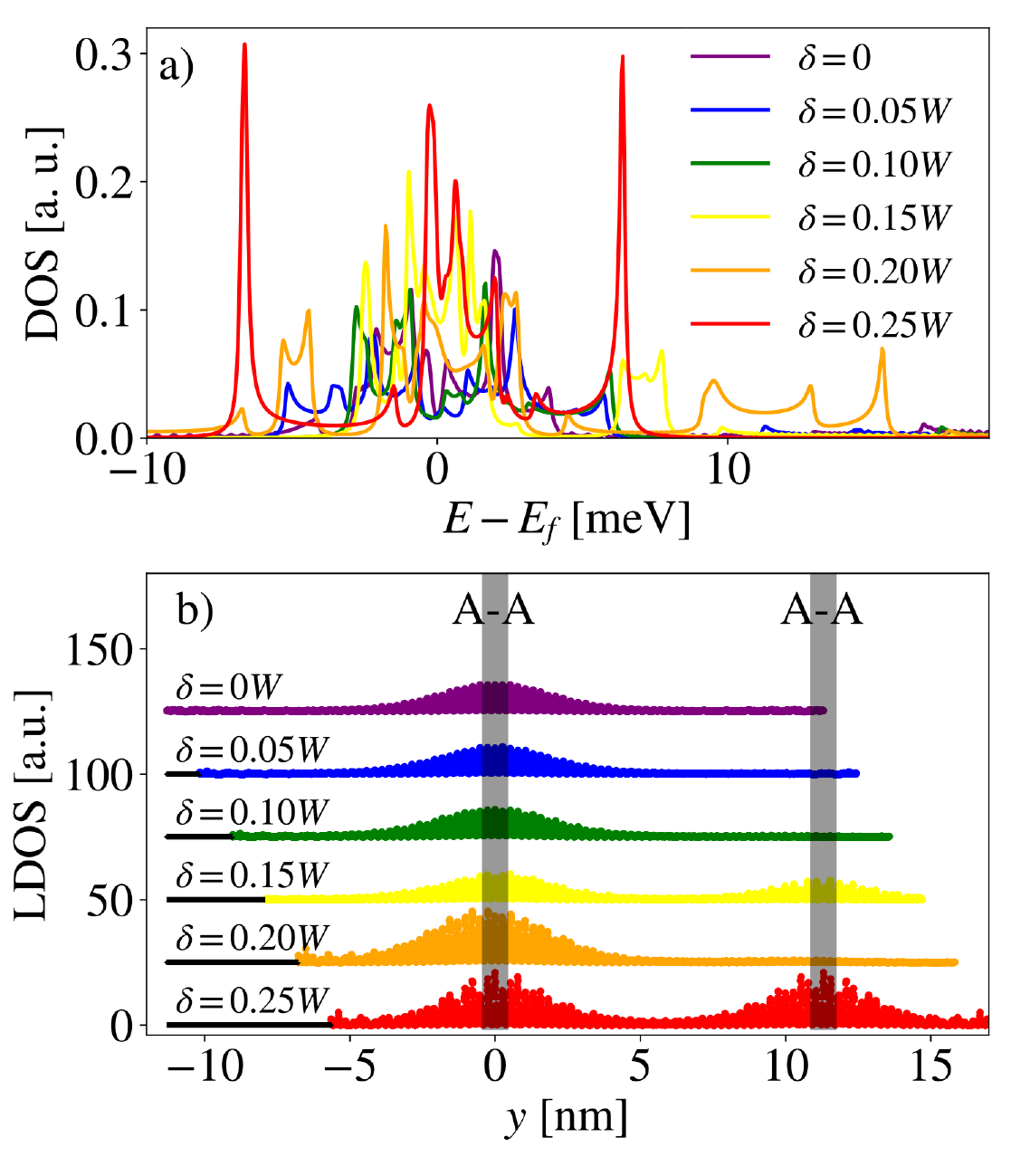}
\caption{a) Density of states near $E_f$ for the ribbons shown in Fig. \ref{fig:ACRibbons}, denoted in different colors for $\delta=0$ (purple), $\delta=0.05W$ (blue), $\delta=0.1W$ (green), $\delta=0.15W$ (yellow), $\delta=0.2W$ (orange) and $\delta=0.25W$ (red). b) Local density of states (LDOS) of the four nearest bands to $E_f$ (blue bands in Fig. \ref{fig:ACRibbons} a)-f)), plotted against the $y$-component of each site. The ribbons are aligned such that the moir\'e centers ($AA$ regions) are located at $y=0$.}
\label{fig:DOS}
\end{center}
\end{figure}

\subsection{Charge localization} \label{sec: Charge Localization}

The effect of an incomplete moir\'e cell at the edge is depicted in Fig.~\ref{fig:DOS}a), where we show the density of states  (DOS) near the $E_f$ for the six borders considered in Fig.~\ref{fig:ACRibbons}. As $\delta$ increases more states move towards $E_f$ due to the completion of the second moir\'e. In Fig.~\ref{fig:DOS}b) the local density of states (LDOS) for the blue bands in Fig.~\ref{fig:ACRibbons}a)-f) is plotted against their $y$-component, with an explicit representation of $\delta$, the black highlighted columns show the location around the $AA$ centers of the moir\'e. As the partial moir\'e  becomes complete, cf. red LDOS in Fig.~\ref{fig:DOS}b), there is a sudden localization of the electrons in the AA sites, in agreement with the experiment in Ref.~\cite{yin2022direct}. 

The previous numerical results suggest that electronic localization plays a paramount role around $E_f$. To further elucidate its role, we perform a localization analysis using the inverse participation ratio (IPR), which is widely used as a measure of wavefunction localization~\cite{bell1970atomic,Thouless1972IPR,Naumis2007MobilityEdge,ShuklaIPR}. It is defined as,
\begin{equation}
\text{IPR}(E)=\sum_i |\psi_{i,k}(E)|^4,
\end{equation}
where the sum runs over the atomic sites. For extended states, the IPR$(E)$ scales as $1/N$, where $N$ is the number of atoms while for localized states goes as IPR$(E) \propto 1/N^{0}$. A clearer description is obtained by using the normalized IPR, which is defined as \cite{Naumis2007MobilityEdge},
\begin{equation}
    \alpha(E)=\frac{\text{log}(\text{IPR}(E))}{\text{log}(N)},
    \label{eq: IPR}
\end{equation}
and thus its range is between -1 and 0.
\begin{figure}
\begin{center}
\includegraphics[width=0.47\textwidth]{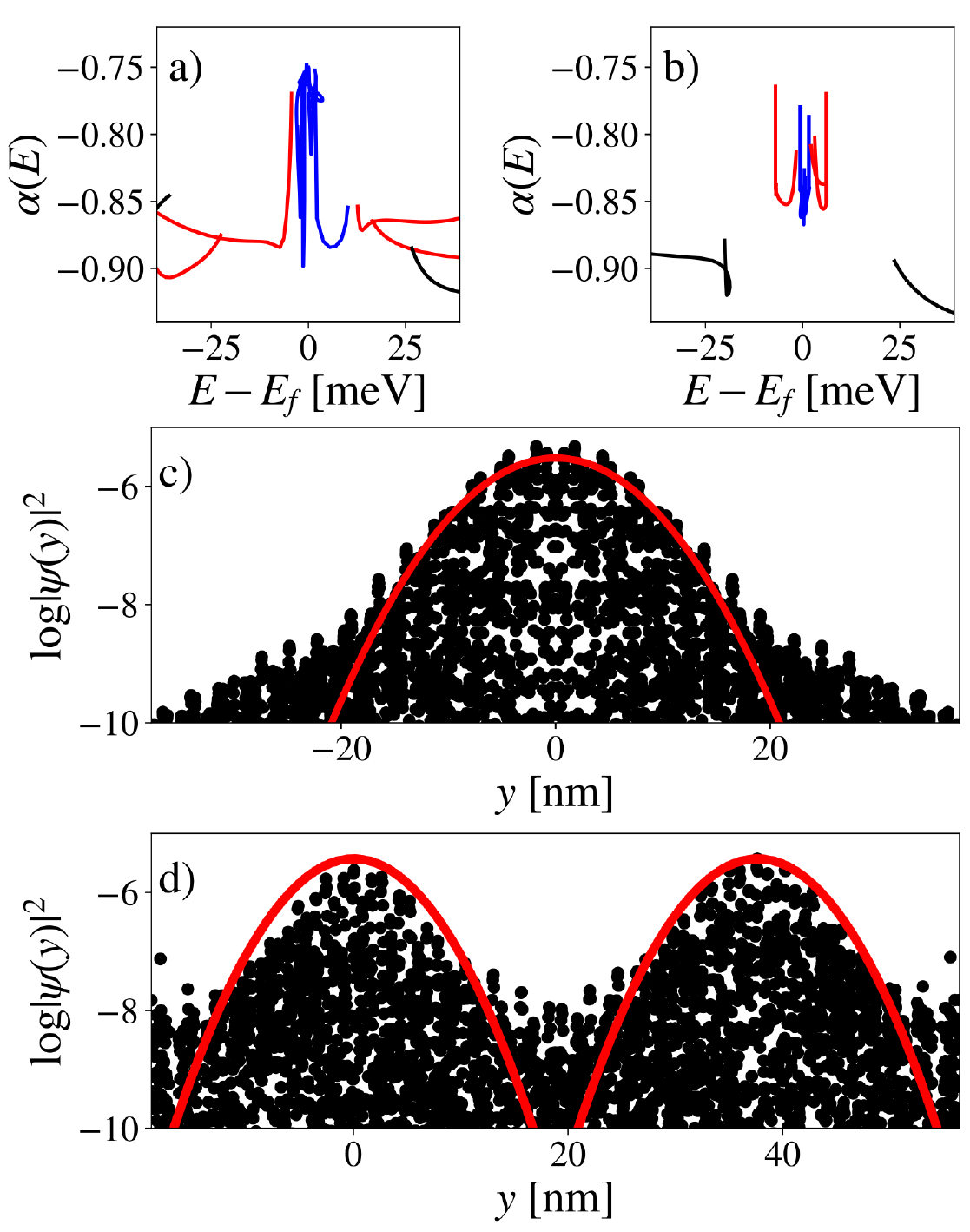}
\caption{Normalized inverse participation ratio $\alpha(E)$ for a) $\delta=0$ and b) $\delta=0.25W$, as expected the peaks occur around $E_f$ where we have the highly localized states at the $AA$ regions. We show in black dots the logarithm of the numerical probability densities for states near $E_f$ against their $y$-coordinate
for c) $\delta=0$ and d) $\delta=0.25W$. The red line corresponds to a Gaussian probability density in which the standard deviation is chosen to match the IPR of the corresponding state. We can see a good agreement around the localization centers.}
\label{fig:IPR}
\end{center}
\end{figure}
In Fig.~\ref{fig:IPR}, we display $\alpha(E)$  for the bands near $E_f$, using the same color code (blue and red) as in Fig.~\ref{fig:ACRibbons}a)-f). Panels a) and b) shows $\alpha(E)$ for $\delta=0$ (a whole moir\'e at the center and moir\'e halves at the edges) and $\delta=0.25W$ (two complete moir\'es), respectively.  As expected, the highest $\alpha(E)$ is located around $E_f$ corresponding to the flat bands (indicated in blue). It is even greater for $\delta=0$ as states are localized around just one moir\'e. The $\alpha(E)$ corresponding to the bands that merge with the flat bands (red bands), change their localization behavior becoming technically localized as flat bands for $\delta=0.25W$. All these results suggest that states do have peculiar localization properties around $E_f$. This is further corroborated in Fig.~\ref{fig:IPR}c)-d), where we display the logarithm of the electron density projected along the nanoribbon cross-section. The density tracks the centers of the moir\'e unitary cells as expected.  

Moreover, a recent analysis of TBG continous models suggest that flat band states are akin to coherent-Landau states~\cite{NavarroLabastida2022Magic,NavarroRMF_2023,FlatBandsElias}. Also, flat band states in strained graphene nanoribbons have been proved to be pseudo-Landau level states akin to soliton states due to topological boundaries in the Jackiw-Rebbi model~\cite{FlatBandsElias}. In both cases, a nearly Gaussian envelope of the flat band states was found. This has also been confirmed in calculations that allows to factorize a lowest-Landau level part in the flat-band wavefunction \cite{Tarnopolsky2019,Popov2020,CanoJennifer2021,2021yardenn}.    
 
 Here a nearly Gaussian-type of localization can be seen in Fig.~\ref{fig:IPR}c)-d) as the maximal log$|\psi(r)|^{2}$ for a given $y$ seem to be limited by downwards parabolas centered at $AA$ sites.  Thus we propose a nearly Gaussian wavefunction with a standard deviation $\sigma$ such that their IPRs match, i.e.,  to equate the IPRs of the numerical results we use a Gaussian wavefunction with $\sigma$ given by, 
\begin{equation}
    \sigma=\sqrt{\frac{1}{2 \pi \text{IPR(E)}}}.
\end{equation}
 For each case we pick a state near $E_f$ and plot log$|\psi(r)|^{2}$  against its corresponding Gaussian fit, indicated by the red curves in Fig. \ref{fig:IPR}c)-d)  for $\delta=0$ and $\delta=0.25W$ respectively. The excellent matching between the red curves and the numerical results near the $AA$ regions indicates a Gaussian behavior in such parts. However, there are clearly fat tails in Fig.~\ref{fig:IPR}c) when compared with a Gaussian indicating the particular shape and properties of flat band functions. These fat tails are responsible for the overlap between different AA regions and highlight the interesting nature of flat-band states. Also, the strong tendency for Gaussian localization in AA regions shown here explains why the nanoribbon width does not affect much the spectrum, instead, is the breaking of the AA regions by edges, in agreement with the experiment in Ref.~\cite{yin2022direct}. The strong localization can also be used to estimate the effects of edges used a toy model based on confinement effects as will be explained in the following section.

\subsection{Toy Model: Moir\'e Potential Well}
\label{sec:toymodel}
\begin{figure}
\begin{center}
\includegraphics[width=0.48\textwidth]{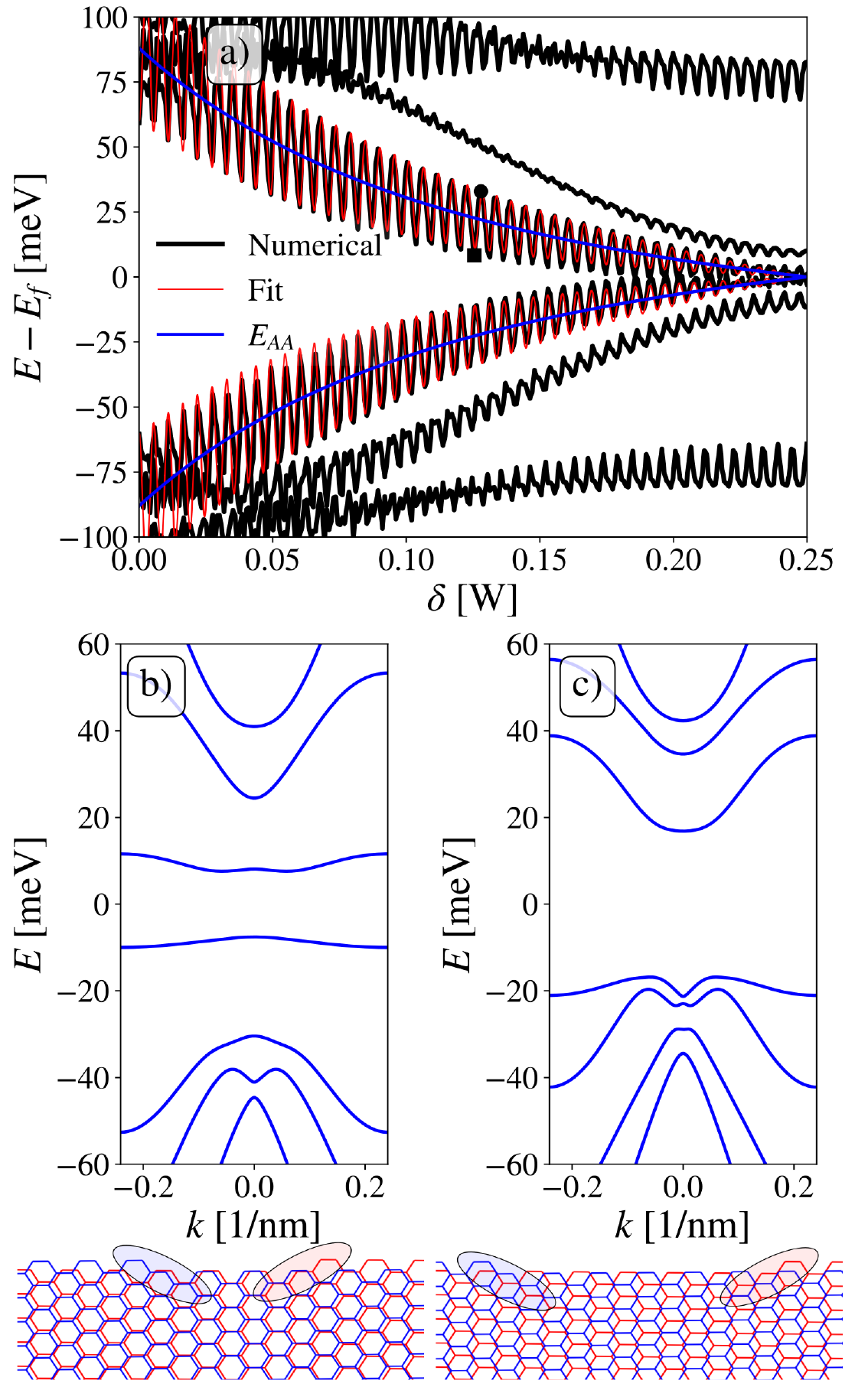}
\caption{a) Energy near $E_f$ of the bands at the Brillouin zone borders for $\theta=1.08^o$ and $W=11.3$ nm (1 moir\'e cell). The black lines show the numerical results, while the red lines is a fit obtained by considering a change in energy akin to a quantum well as a function of the well's width (blue line), plus an oscillating perturbation originated by the \textit{zigzag} borders. The square and circle markers shows respectively the local minimum and maximum for similar values of $\delta$. The low-energy spectrum and the edges of the ribbon is shown for both cases indicated by the markers. In b) the bands are nearer to $E_f$ and the \textit{zigzag} edges are at $AA$ stacking sites, while in c) the bands are further away from $E_f$ and the \textit{zigzag} edges are at $AB$ sites.}
\label{fig:Fit}
\end{center}
\end{figure}

As shown in Fig. \ref{fig:MoireEdges}, the lowering of the bands energy towards $E_f$ as $\delta$ increases is composed of two effects. The first one is the global tendency associated to the confinement. This effect can be explained within the moir\'e quantum well picture.  As confined states mainly localized in a quantum well have an energy dependence of $E(L) \propto 1/L^2$, where $L$ is the length of the well. Thus, when we increase the parameter $\delta$, the size of the well is increased and this lowers the energy due to less quantum confinement. We then propose that the energy of the bands localized at the partial moir\'e have the following $\delta$ dependence,
\begin{equation}
    E_{AA}(\delta)=E_0+\frac{AW^2}{(\delta-\delta_0)^2},
    \label{Eq:Confinement} 
\end{equation}
where $A$ and $E_0$ are coefficients to be determined, and $\delta_0=-0.25W$ is the value  where the partial moir\'e cell vanishes completely. With the numerical data we can use the conditions at $E_{AA}(\delta=0)=0$ and $E_{AA}(\delta=0.25W)=88$ meV to obtain the coefficients in Eq.~(\ref{Eq:Confinement}), which give $A=7.33$ meV and $E_0=-29.33$ meV. In Fig. \ref{fig:Fit}a)  we plot the curve given by Eq.~(\ref{Eq:Confinement}) using these parameters alongside the numerical results.  Clearly, Eq.~(\ref{Eq:Confinement}) correctly describes the energy lowering without the oscillating component.
As we stated in the previous section, the oscillation wavelength is invariant prior to the rescaling, and this indicates that its origin comes from the graphene lattice. One thing to notice is that due to the rotation angle, the graphene lattice is not aligned with the ribbon's periodicity, thus there is a discontinuity in the \textit{armchair} edge at which a small \textit{zigzag} border appears. We found that the values of $\delta$ at which the oscillation has its local maximums occur when this \textit{zigzag} border lies in the $AB$ regions, while the local minimums occur when it is in the $AA$ regions. In Fig.~\ref{fig:Fit}b)-c) we display the bands and corresponding edges for two values of $\delta$ indicated by an square and circle in Fig.~\ref{fig:Fit}a), respectively. Both ribbons have moir\'e cells of roughly the same size.

Therefore, we recognize that the position of the \textit{zigzag} border moves periodically as a function of $\delta$, and its periodicity corresponds to $\lambda$, the wavelength of the small oscillation in the spectrum. Thus, we propose that the \textit{zigzag} edges are responsible for the oscillating perturbation to the spectrum, we model them as,
\begin{equation}
    \Delta E(\delta)=B(e^{-(\delta+\delta_0)/\gamma}-1)\text{cos}\left(\frac{2\pi \delta}{\lambda}\right).
\end{equation}
Here, $B$ is the amplitude of the perturbation, and $\gamma$ modulates an exponential decay that we included to control the strength of the perturbation. As the moir\'e becomes mostly complete, the spectrum is predominantly governed by the moir\'e well. The energy of these states, mainly localized at the partial moir\'e, as a function of $\delta$ is then,
\begin{equation}
    E(\delta)=E_{AA}(\delta)+\Delta E(\delta).
    \label{Eq:Fit}
\end{equation}
We found that $B=-32$ meV and $\gamma=0.4W$ fit the numerical results as shown in Fig.~\ref{fig:Fit}a). Interestingly, the large scale oscillations that we found are consistent with those reported in Ref.~\cite{Fujimoto2021pumping}.  However, it seems that the impact of the small oscillations was not been fully considered. This perturbation of the moir\'e flat bands has not been reported to our knowledge, and the exact form of the coupling between moir\'e and edge states still requires further research~\cite{fleischmann2018moire,Wang2023ZigZagNano}. Furthermore, in a real sample, regions with \textit{zigzag} edges this coupling will inevitably emerge. Therefore, to complement this work, in the following section, we provide a similar analysis to the one shown in Fig. \ref{fig:ACRibbons} but for \textit{zigzag}-terminated nanoribbons.

\section{Zigzag terminated TBG nanoribbons} \label{sec:zz}
\begin{figure}
\begin{center}
\includegraphics[width=0.47\textwidth]{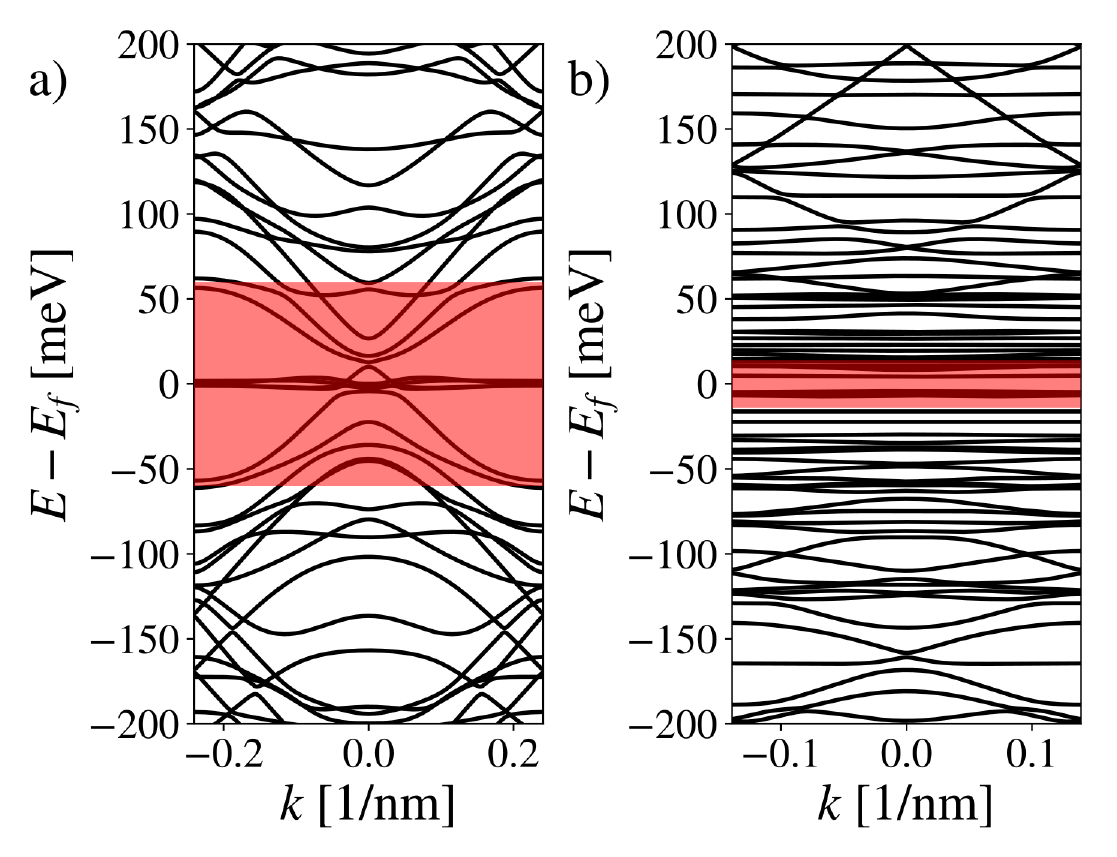}
\caption{Energy spectrum for TBG nanoribbons with a) {\it armchair} termination and a width $W_{AC}=22.6$ nm and b) {\it zigzag} termination and a width $W_{ZZ}=13$ nm. Both ribbons are considered with $\delta=0$ such that each ribbons has a complete moir\'e at the center and moir\'e halves at the borders. The red highlighted areas in a) and b) show the energy windows at which the spectrum is shown in Fig.~\ref{fig:ACRibbons} and Figs.~\ref{fig:ZZRibbons} and \ref{fig:ZZRibbons2} respectively.        }
\label{fig:Window}
\end{center}
\end{figure}

In this section, we investigate the scenario of {\it zigzag} terminated TBG nanoribbons. The ribbon's unit cell is illustrated in Fig.~\ref{fig:system}c). Unlike the {\it armchair}  case in the previous section, the {\it zigzag} configuration can be achieved through a translation of a single shaded cell along the $y$-direction. Figure~\ref{fig:Window} depicts a comparison of the band structure for the nanoribbons of the two considered configurations. The energy bands for the {\it zigzag} configuration in Fig.~\ref{fig:Window}b) indicate the presence of low-dispersive states even at high energies. In contrast, for the {\it armchair} configuration in Fig.~\ref{fig:Window}a), apart from the localized states in the middle of the spectra, there are fewer low-dispersive energy states. 

In the previous section, we found that the states in the small regions with {\it zigzag} edges, as shown in Fig.~\ref{fig:Fit}b)-c), are responsible for the small oscillations of the moir\'e bands, as seen in Fig.~\ref{fig:Fit}a). However, the perturbation at the moiré scale is minor and the spectrum is dominated by bulk bands, in this situation, the spectrum for the \textit{armchair} termination clearly shows the flat-band structure of TBG. In contrast, for a {\it zigzag} nanoribbon, the boundaries generate several edge states that significantly impact the moiré spectrum, as depicted in Fig.~\ref{fig:Window}b). 

\begin{figure}
\begin{center}
\includegraphics[width=0.49\textwidth]{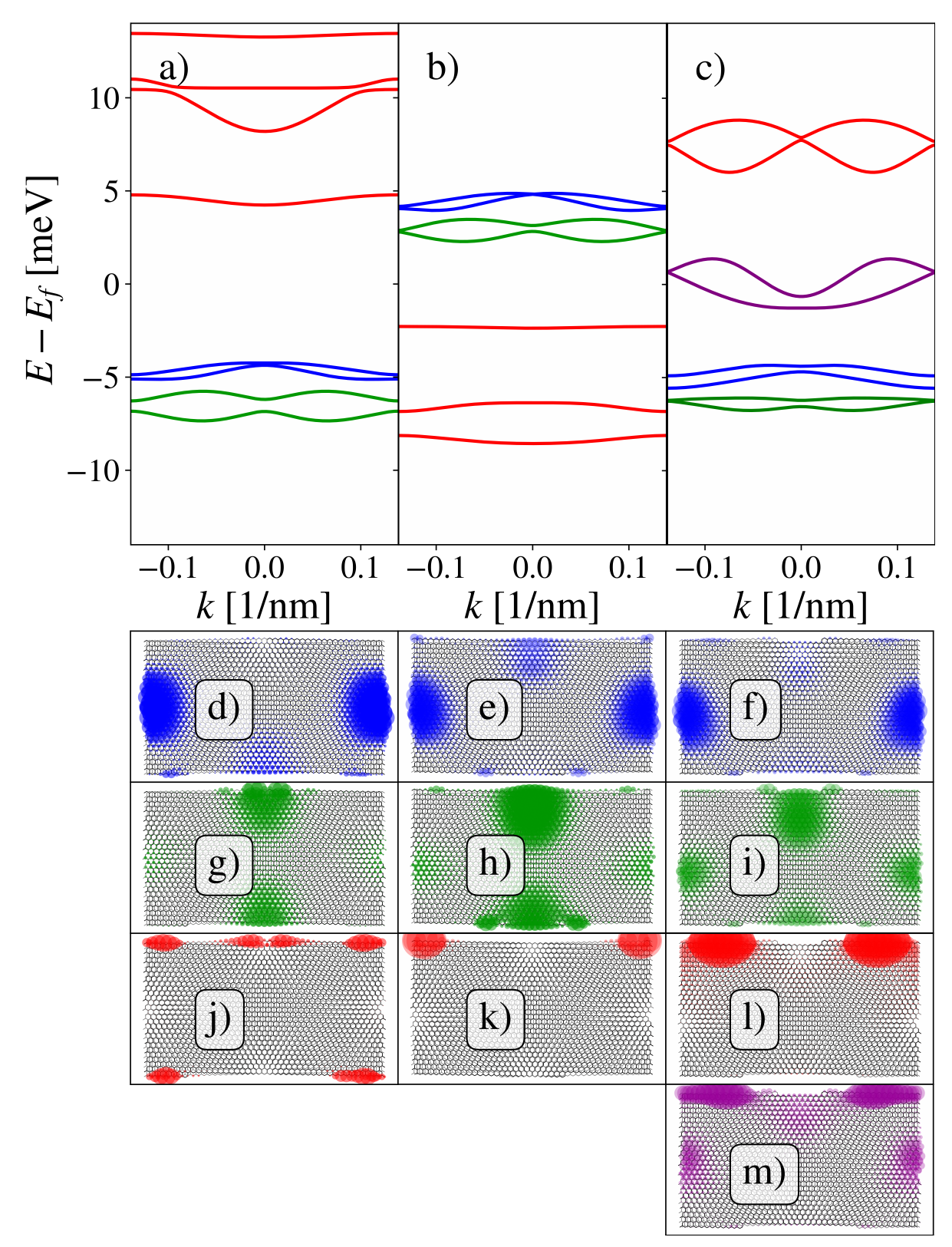}
\caption{Low energy band structure for {\it zigzag terminated} TBG nanoribbons simulating a twist angle $\theta=1.08^o$ with constant width $W=13$ nm but for different moir\'e terminations obtained by translating the ribbon unit cell by a) $\delta=0$, b) $\delta=0.05W$ and c) $\delta=0.1W$. We color the bands near $E_f$ according to their localization, blue for states localized near the $AA$ regions of the mostly complete moir\'e cells near the center of the ribbon, green for the states localized near the $AA$ regions of the incomplete moir\'e cells near the ribbon edges, red for edge states and purple for states localized at both $AA$ regions and edges. Below each set of bands we plot the corresponding charge density indicated with the same color code.}
\label{fig:ZZRibbons}
\end{center}
\end{figure}

\begin{figure}
\begin{center}
\includegraphics[width=0.47\textwidth]{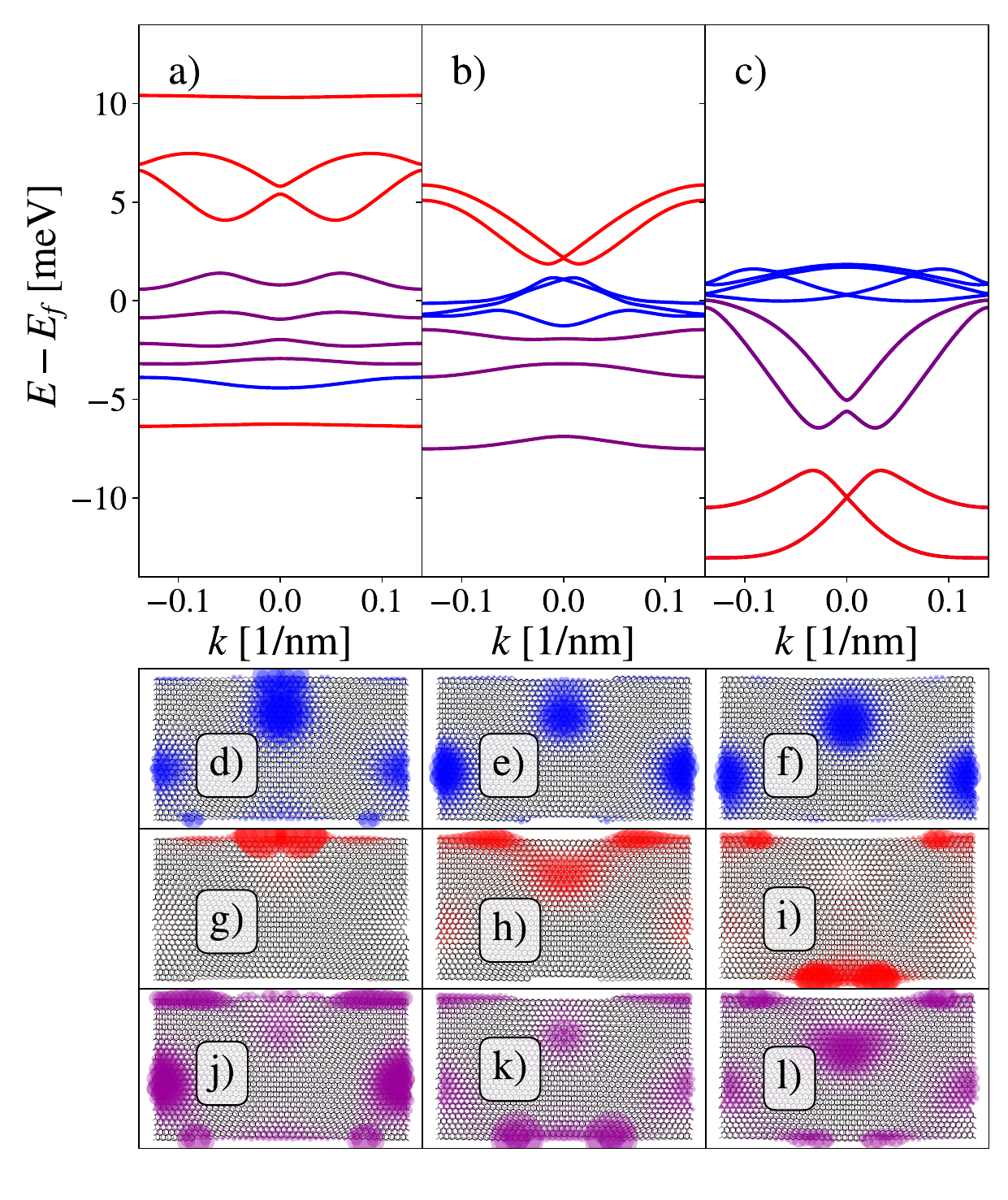}
\caption{Low energy band structure for {\it zigzag terminated} TBG nanoribbons simulating a twist angle $\theta=1.08^o$ with constant width $W=13$ nm, we consider different moir\'e terminations by translating the unit cell of the ribbon by a) $\delta=0.15W$, b) $\delta=0.2W$ and c) $\delta=0.25W$. We color the bands near $E_f$ according to their localization, blue for states localized near the $AA$ regions, red for edge states and purple for states localized at both $AA$ regions and edges. Below each set of bands we plot the total charge density in the bands of the respective color.}
\label{fig:ZZRibbons2}
\end{center}
\end{figure}

In Fig.~\ref{fig:ZZRibbons} and Fig.~\ref{fig:ZZRibbons2}, we characterize different moir\'e terminations by modifying the sliding parameter $\delta$. We ensure that there is always at least one complete moir\'e within the ribbon. Our focus is on the energy range marked by the red shaded region in Fig.~\ref{fig:Window}b). It is worth noting that this energy range is narrower than that of the armchair case due to the denser spectrum near the Fermi energy. Figures~\ref{fig:ZZRibbons}a)-c) and \ref{fig:ZZRibbons2}a)-c) depicts the band structure for different sliding values. We classify the bands according to their localization: blue for bands mostly localized at the $AA$ regions, red for edge states, green for (mixed) states that are a mixture between AA localized and moire edge states, and purple for those that are a mixture between AA localized and edge states. We notice the difference between moir\'e edge state and edge state, while the first one is an state mainly localized in an incomplete moir\'e along the edge (green), the second one is an state mainly localized on the zigzag sites along the boundary (red). The evolution of the states as the sliding parameter is modified is not as straightforward as in the {\it armchair} case, as each edge and bulk state has a different energy due to the border geometry. This complexity makes it challenging to differentiate them solely based on their energy and reveals the impact of the edge microscopic boundary in the electronic spectrum. 

In Fig.~\ref{fig:ZZRibbons}d)-m) and Fig.~\ref{fig:ZZRibbons2}d)-l) we show the total charge density for the corresponding band structures. This helps us distinguish the localization of states in the different energy bands. Below each band structure we show the charge density for the bulk, edge, and mixed bands. It is important to note that this classification is not strict; we are using it solely to differentiate the states based on their maximum localization. Under this criterion, as shown in Fig.~\ref{fig:ZZRibbons} and Fig.~\ref{fig:ZZRibbons2}, is clear that the bulk moire bands (blue), are mainly localized in the AA centers, while the edge states (red), mainly appear at the $AB$ regions. These edge states are reminiscent of the monolayer {\it zigzag} configuration~\cite{Fujita1996Peculiar,Nakada1996EdgeState} and their localization in the AB sites is in agreement with experiments in moir\'e structures of graphite~\cite{Morell2014TBGnanorribbons,Morell2015EdgeGraphite}.

\begin{figure}
\begin{center}
\includegraphics[width=0.48\textwidth]{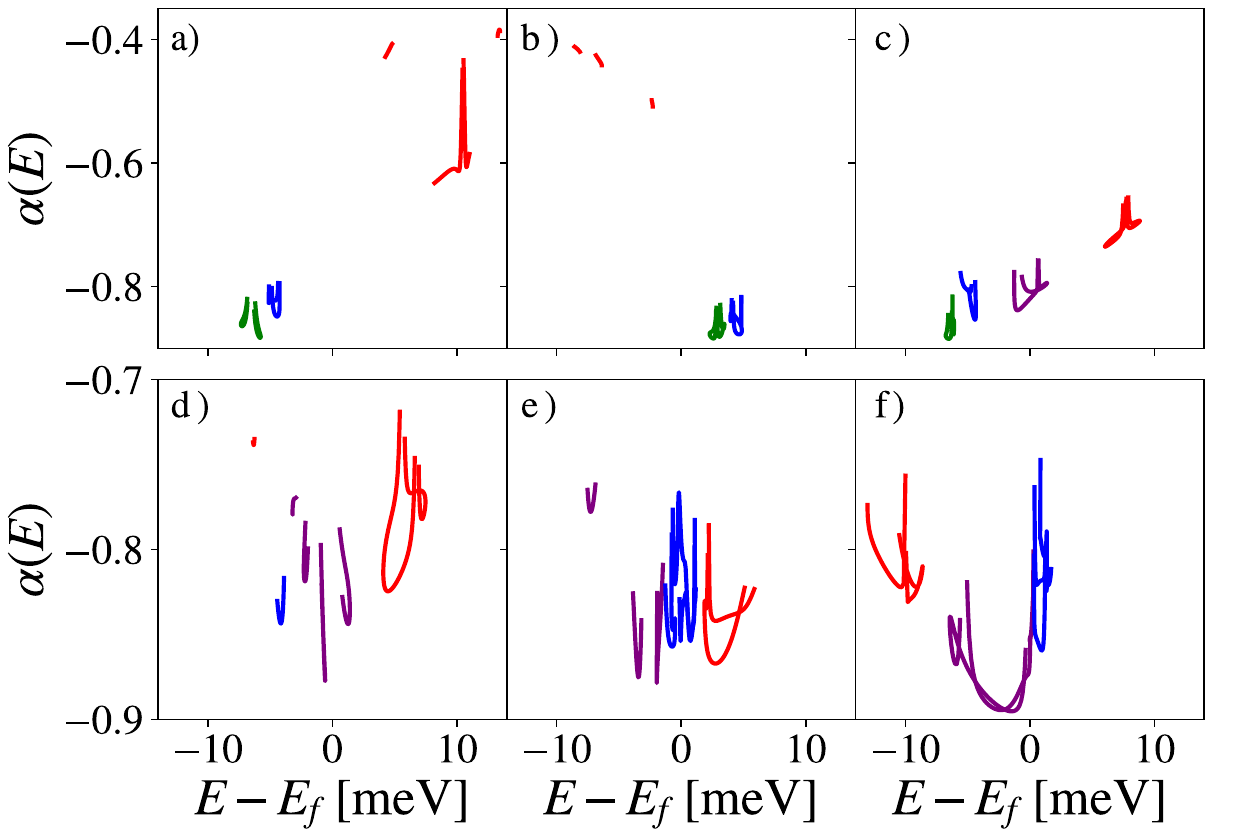}
\caption{Normalized inverse participation ratio $\alpha(E)$. Panels a)-c) correspond to the bands shown in Fig. \ref{fig:ZZRibbons} and d)-f) for the bands in Fig. \ref{fig:ZZRibbons2}. The color code is the same as in Fig. \ref{fig:ZZRibbons}-\ref{fig:ZZRibbons2}. As expected, the most localized states usually correspond to edge states (red) as they are distributed between a small number of atoms, especially in a) and b). Notice the different scales used between a)-c) and d)-f). On the contrary case, states that are distributed between the $AA$ regions and the edges (purple) usually have the lowest value of $\alpha(E)$.} 
\label{fig:ZZIPR}
\end{center}
\end{figure}

To characterize the different bulk and edge charge contributions, we now perform a localization analysis. Figure~\ref{fig:ZZIPR} shows the parameter $\alpha(E)$ in Eq.~\ref{eq: IPR} around $E_f$. In the top panels, the mixed states (green) mostly occupy the incomplete moir\'e cells with a small contribution from the AA centers, cf. Fig~\ref{fig:ZZRibbons}g)-i), resulting in a low value of $\alpha(E)$. The values of $\alpha(E)$ for the blue states are just slightly above the green ones, as they are mostly contained within the complete moir\'e cells. Both blue and green states have a similar value of their $\alpha(E)$ to those of the \textit{armchair} case as they arise from the moir\'e pattern. However, the red edge states show a clear difference in magnitude as they are distributed in a smaller number of atomic sites. In the bottom panels, Fig.~\ref{fig:ZZIPR}d)-f), the edge states are not as localized as in the previous case, most of them are partially hybridized with moir\'e states (notice the difference in the scales). The purple states are highly hybridized moir\'e and edge states, and thus have the lowest values of $\alpha(E)$ in most cases. The blue states on the other hand, have their maximum value of $\alpha(E)$ for $\delta=0.25W$, as this is the point where the two different moir\'e cells present are mostly complete, thus the wavefunction becomes maximally localized within the $AA$ regions.

%\EA{We now perform a localization analysis for the states shown in Figs.~\ref{fig:ZZRibbons} and \ref{fig:ZZRibbons2}. The parameter $\alpha(E)$, cf. Eq.~\ref{eq: IPR}, near $E_f$ is shown in Fig.~\ref{fig:ZZIPR}. In the top panels, Fig.~\ref{fig:ZZIPR} a)-c), the green states are mostly localized at the incomplete moir\'e cells, although they are still distributed to a lesser degree in the complete moir\'e cells, this results in a low value of $\alpha(E)$. The values of $\alpha(E)$ for the blue states are just slightly above the green ones, as they are mostly contained within the complete moir\'e cells. Both these states have a similar value of their $\alpha(E)$ to those of the \textit{armchair} case as they arise from the moir\'e pattern. However, the red states corresponding to edge states show a clear difference in magnitude as they are distributed in a smaller number of atomic sites. In Fig. \ref{fig:ZZIPR} d)-f) the edge states are not as localized as in the previous case, most of them are partially hybridized with moir\'e states (notice the difference in the scales). The purple states are highly hybridized states of moir\'e and edge states, and thus have the lowest values of $\alpha(E)$ in most cases. The blue states on the other hand have their maximum value of $\alpha(E)$ for $\delta=0.25W$, as this is the point where the two different moir\'e cells present are mostly complete, thus the wavefunction becomes maximally localized within the $AA$ regions. }

A recent experiment~\cite{yin2022direct} has reported that the low energy flat band can exist as long as the moir\'e cell remains complete. Tunneling spectra revealed moir\'e AA spots when the tip was inside the sample. However, an absence of localization due to an incomplete moir\'e was observed at the boundary. As previously described for an armchair configuration, we have found that the nearly Gaussian localization of the moir\'e bulk states survives if the moir\'e is complete, cf. Fig.~\ref{fig:IPR}c)-d) and is suppressed if the moir\'e is incomplete, cf. Fig.~\ref{fig:DOS}b). This effect is also present in a {\it zigzag} geometry, where the maximum localization is when the moir\'e is completed, cf. Fig.~\ref{fig:ZZIPR}f). Interestingly, we found that if the moir\'e is incomplete, there is an strong hybridization with less localized edge states.  

In the experiment described in Ref.~\cite{yin2022direct}, the edge orientation predominantly features a zigzag termination. Their results indicated a breakdown of the localized flat bands for an incomplete moir\'e, and they also reported the presence of edge states in their LDOS maps, aligning with our findings in Fig.~\ref{fig:ZZRibbons}. These edge states, represented by the red and green states in Fig.~\ref{fig:ZZRibbons}, manifest as residual LDOS peaks in incomplete moires and rapidly decay with distance from the edges. The AA localized states only appear in complete moires, in agreement with the blue states in Fig.~\ref{fig:ZZRibbons2}. Given that the bulk localization is linked to the TBG flat bands, our results demonstrate their sensitivity near the boundaries. Furthermore, for ribbons with widths spanning multiple moir\'e cells, the flat bands in the bulk remain unperturbed as we modify the borders. We believe that our work comprehensively captures the essential characteristics of both bulk and edge states in TBG nanoribbons and it offers a detailed explanation for the experimental results in Ref.~\cite{yin2022direct}.

\section{Conclusions}
\label{sec:conclusion}

In this work, we have examined the effects of edges on the electronic properties of twisted bilayer graphene nanoribbons. Using a tight-binding model, we characterized the edge and bulk states in nanoribbons with \textit{zigzag} and \textit{armchair} edge terminations. We found that the flat bands in ribbons with widths spanning multiple moir\'e cells remain insensitive to edge modifications. However, near the boundaries, flat band localization is sensitive to the edge termination. We observed that the flat band exhibits nearly Gaussian localization, which is suppressed if the moir\'e is incomplete. Additionally, depending on the edge orientation, bulk states can hybridize with the edge states. 

In \textit{armchair}-terminated nanoribbons, the low-energy spectrum is determined by the completeness of the moir\'e cells at the ribbon's edge, transitioning from dispersive to flat bands as the moir\'e cell becomes complete. This behavior aligns with a moir\'e quantum well picture. There is also a perturbation in the band's energy due to the appearance of small regions with zigzag termination. Depending on whether these zigzag borders appear on AA or AB stacking regions, the band energy is either lowered or raised, while the main Gaussian-like localization in the AA regions remains. These zigzag regions will inevitably appear due to the mismatch between the ribbon alignment and the armchair direction.

In \textit{zigzag}-terminated nanoribbons, there is a more pronounced impact on the low-energy spectrum due to the edge states, as they can become hybridized with the moir\'e states. However, some states remain predominantly localized within the AA regions. The influence of the edge states is expected to be diminished in real samples as translational symmetry is reduced, resulting in a mixture of armchair and zigzag boundaries. As in the \textit{armchair} case, the behavior should primarily be described by the bands localized within the AA regions. Consequently, regardless of the edge termination, we expect AA localization when the moir\'e is complete and delocalization when it is not. Our findings provide an explanation for the breakdown of flat bands and the coexistence of bulk and edge states in recent experiments with twisted bilayer graphene nanoribbons~\cite{yin2022direct}

\section{Acknowledgments}
We thank H\'{e}ctor Sainz-Cruz for fruitful discussions. This work was supported by UNAM DGAPA PAPIIT IN102620 (E.A. and G.G.N.), CONAHCyT project 1564464 (E.A.,G.G.N.).  IMDEA Nanociencia acknowledges support from the \textquotedblleft Severo Ochoa\textquotedblright ~Programme for Centres of Excellence in R\&D (Grant No. SEV-2016-0686). P.A.P and F.G. acknowledge funding from the European Commission, within the Graphene Flagship, Core 3, grant number 881603 and from grants NMAT2D (Comunidad de Madrid, Spain), SprQuMat and (MAD2D-CM)-MRR MATERIALES AVANZADOS-IMDEA-NC.

\bibliographystyle{unsrt}
\bibliography{refs.bib}
\end{document}